\documentclass[prl,twocolumn,showpacs,amsmath,amssymb,floatfix]{revtex4}
\usepackage{graphicx}
\usepackage{dcolumn}
 \pdfoutput=1
\usepackage{bm}
\usepackage{amsmath}
\usepackage{latexsym}
\usepackage{amsfonts}
\usepackage{amssymb}
\usepackage[rightcaption]{sidecap}
\usepackage{float}
\makeatletter

\newcommand{\Rmnum}[1]{\expandafter\@slowromancap\romannumeral  #1@}
\makeatother

\begin{document}
\title{Successes and Failures of Kadanoff-Baym Dynamics in Hubbard Nanoclusters}
\date{\today}
\author{M. Puig von Friesen}
\author{C. Verdozzi}
\author{C.-O. Almbladh}
\affiliation{Mathematical Physics and European Theoretical Spectroscopy Facility (ETSF), Lund University, 22100  Lund, Sweden}
\begin{abstract}
We study the non-equilibrium dynamics of small, strongly correlated clusters, described by a Hubbard Hamiltonian, by propagating in time the Kadanoff-Baym equations within the Hartree-Fock, 2$^{nd}$ Born, GW and T-matrix approximations. We compare the results to exact numerical solutions. We find that the T-matrix is overall superior to the other approximations, and is in good agreement with the exact results in the low-density regime. In the long time limit, the many-body approximations attain an unphysical steady state which we attribute to the implicit inclusion of infinite order diagrams in a few-body system.
\end{abstract}
\maketitle
Introduced almost fifty years ago, the Kadanoff-Baym equations (KBE) \cite{KB-book,Keldysh} are a cornerstone of a microscopic theory of non-equilibrium quantum processes \cite{Danielewicz, Bonitzbook}. In recent years, the KBE have received growing attention by the strongly correlated systems (SCS) and quantum transport (QT) communities, due to increased computational capabilities 
\cite{bonitz1,Jauho,Stefanucci04,Robert,Galperin,Thygesen1,Olevano, RobertHubb, Freericks}. 
Non-perturbative treatments of SCS with the KBE are beginning to appear, albeit specialized
to spatially uniform fields \cite{Freericks}. At present, for general time-dependent (TD) {\it and} inhomogeneous fields, the only available treatments of the KBE for SCS rely on many-body perturbation theory (MBPT), where one systematically builds approximations for the one-particle Greens function, $G$. The latter is the key quantity in the KBE. Most work with the KBE has been so far for model systems but  {\it ab initio} studies based on MBPT+KBE are beginning to appear \cite{Robert,Thygesen1,Olevano}. 

Since MBPT+KBE currently appears to be the only route to SCS for general perturbations,  it is rather surprising that a rigorous assessment of the applicability of the approach has not yet been made. In this work, we perform such an evaluation, by comparing different MBPT schemes for the KBE against exact 
solutions for small, isolated clusters exposed to TD inhomogeneous perturbations. Clearly, in  
finite closed systems, one cannot address important issues such as the onset of a steady state regime, 
or the role of dissipation: This requires coupling the clusters to an environment as done, for example, 
in \cite{Thygesen1} for the stationary limit, or in \cite{RobertHubb} for the TD case.  
However, as shown here, studying small isolated clusters can reveal
crucial and unexpected finite size effects in the nature of the KBE solutions.

Our clusters are small 1D chains with Hubbard interactions. For the MBPT+KBE,  we 
considered four conserving \cite{KB-book} many-body approximations (MBAs):  Hartree-Fock, $2^{nd}$ Born,
GW \cite{Hedin} and T-matrix \cite{TMA, BonitzTMA} (HFA, BA, GWA and TMA, respectively).
We have also used a spin dependent GWA (SGWA) \cite{Thygesen1,Thygesen2} to mitigate spurious self-interaction effects
\cite{Pina,Rex}. 

Another common approach to microscopic TD phenomena is Time Dependent Density Functional Theory (TDDFT)  \cite{Hardy}.
A well established connection exists between TDDFT 
and MBPT on the Keldysh contour \cite{TDDFT}.
Here, to highlight this link, we will obtain MBPT-based 
exchange-correlation (xc) potentials via TD  reverse engineering \cite{VerdozziPRL}, using the TD 
densities from the KBE.

The main findings of this paper are (i) for Hubbard-like interactions, the TMA performs very 
well at low densities, both for ground-state and time dynamics, and is in general superior to the other MBAs at all electron densities. The difference in  performance among the MBAs increases at larger interaction strength $U$ (ii) the SGWA performs better than the GWA; however, the SGWA breaks down when $U$ exceeds a critical value iii) a comparison of exact vs MBPT-based xc potentials for TDDFT
shows trends similar to those for the densities iv) during 
the time evolution, the KBE develop an unphysical steady state solution. This 
is a central result of the present work, and we argue that this problem occurs in general, whenever
MBPT is applied to finite systems, and self energies based upon infinite partial summations are used.

\noindent \textit{Model system.} The Hamiltonian of our open-ended Hubbard chains is (we set
the onsite energies equal to 0)
\begin{equation}
H\!\!= -V\!\!\!\sum_{\left\langle RR'\right\rangle \sigma}\!\!\!\!a_{R\sigma}^{\dagger}a_{R'\sigma}+
U\sum_{R}\!\hat{n}_{R\uparrow}\hat{n}_{R\downarrow}+\!\!\sum_{R\:\sigma}w_R\left(t\right) \hat{n}_{R\sigma}
\end{equation}
Here $\hat{n}_{R\sigma}=a_{R\sigma}^{\dagger}a_{R\sigma}$, $\sigma=\uparrow,\downarrow$, and $\left\langle RR'\right\rangle $ denotes nearest neighbor sites.  The hopping parameter $V=1$ and $w\left(t\right)$ is a local external field which can be of any shape in time $t$ and space. $U$ and $w\left(t\right)$ are given in
units of $V$. We consider chains of lengths $L=2,6$ and $N_{e}=2,6$
number of electrons; we take spin-up and -down
electrons equal in number, $N_\uparrow=N_\downarrow$; this holds at all times, since
$H$ has no spin-flip terms. Henceforth, $n=N_\uparrow/L$. 

\begin{figure}[t]
\begin{center}
\includegraphics[width=5.8cm, clip=true]{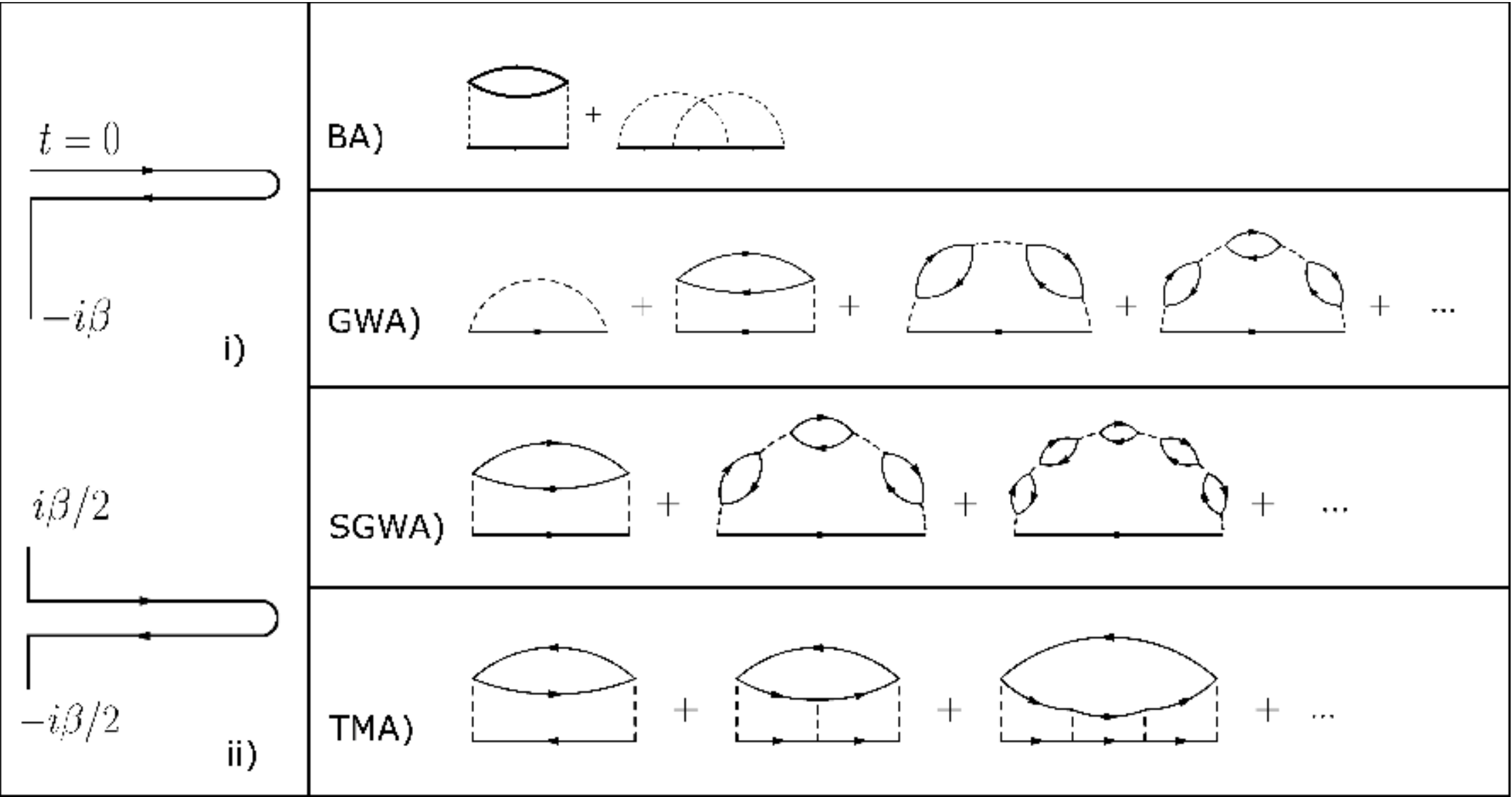}\\
\end{center}
\caption{ Contours and approximate self-energies}
\label{method}
\end{figure}
\noindent \textit{Kadanoff-Baym equations.} 
We find the non-equilibrium one-particle Greens function $G\left(t_{1},t_{2}\right)$
by solving its equation of motion $\left(i\partial_{t_{1}}-h\left(t_{1}\right)\right)G\left(t_{1},t_{2}\right)=\int_{\gamma}\Sigma\left(t_{1},t\right)G\left(t,t_{2}\right)dt$ \cite{KB-book,TDDFT}, and the one corresponding to $t_{2}$. Here $h$ is the single particle Hamiltonian, $\Sigma$ is the self 
energy and $\gamma$ represents the Keldysh contour \cite{Keldysh}. Instead of using the more conventional contour of Fig. \ref{method}i we use an equivalent one (Fig. \ref{method}ii) \cite{Bonitzbook}, numerically more stable and with an analytical limit when the temperature $T\equiv \beta^{-1} \rightarrow0$. The KBE for $t_1$ then become
\begin{eqnarray}
&\hbox{}&\!\!\!\!\left(i\partial_{t_{1}}-h\left(t_{1}\right)\right)G^{\lessgtr}\left(t_{1},t_{2}\right)=\nonumber\\
&\hbox{}&\!\!\!\!\int_{0}^{t_{1}}d\overline{t}\left[\Sigma^{R}\left(t_{1},\overline{t}\right)G^{\lessgtr}\left(\overline{t},t_{2}\right)+\Sigma^{\lessgtr}\left(t_{1},\overline{t}\right)G^{A}\left(\overline{t},t_{2}\right)\right]\nonumber\\
&+&\!\!\!\!\frac{1}{i}\int_{0}^{\beta/2}d\overline{\tau}\Sigma^{<}\left(t_{1},-i\overline{\tau}\right)G^{>}\left(-i\overline{\tau},t_{2}\right)\nonumber\\
&+&\!\!\!\!\frac{1}{i}\int_{0}^{\beta/2}d\overline{\tau}\Sigma^{>}\left(t_{1},i\overline{\tau}\right)G^{<}\left(i\overline{\tau},t_{2}\right)
\label{KB}
\end{eqnarray}
Here $\lessgtr$ and $R/A$ have the usual meaning \cite{KB-book,Keldysh}. We work at $T=0$. The initial state is the correlated \cite{Danielewicz} ground state, obtained by solving the Dyson equation $G=G_{0}+G_{0}\Sigma[G] G$  self consistently, with $(\epsilon-h)G_0=1$.
Eq. (\ref{KB}) is solved by time propagation using a predictor-corrector method similar to that presented in \cite{Kohler}.  Conservation laws and time reversal symmetry are obeyed up to arbitrary accuracy. In a finite system, exact and approximate spectral functions are meromorphic. For example, for the Greens function, $G_{RR'}\left(\epsilon\right)=\sum_{j} A_{RR'}^{j}(\epsilon-a_{j})^{-1}$, 
where $a_{j}$ are pole positions and $A_{RR'}^{j}$ are residue matrices
in the single particle orbital representation. This representation is very convenient as convolutions 
become simple matrix products \cite{MCCV87}. 

\noindent  \textit{Many-body approximations (MBAs).} In the Hubbard model, the interaction can be treated either as spin-dependent, $U\sum_{R}n_{R\uparrow}n_{R\downarrow}$ or as spin-independent, $\frac{1}{2}U\sum_{R\sigma\sigma'}a_{R\sigma}^{\dagger}a_{R\sigma'}^{\dagger}a_{R\sigma'}a_{R\sigma}$.
These two ways are evidently equivalent in any order by order expansion (Fig. \ref{method}) such as the HFA or  BA. In approximations, however, this equivalence may be lost. For example, in spin-independent GWA, $\Sigma\left(12\right)=G\left(12\right)W\left(12\right)$, where $W=U+UPW$ and
 $P\left(12\right)=G\left(12\right)G\left(21\right)$. In spin-dependent GW (SGWA), instead, $W=UPU+\left(UP\right)^{2}W$. The TMA, where $\Sigma\left(12\right)=U^2G\left(21\right)T\left(12\right)$, is treated only as spin-dependent, with $T=\phi-U\phi T$ and $\phi\left(12\right)=G\left(12\right)G\left(12\right)$.

\noindent \textit{Exchange-correlation potential.} From the TD densities we obtained via reverse engineering
the corresponding $v_{KS}=v_{H}+v_{xc}$, $v_{H}$ and $v_{xc}$ being
the Hartree and the xc potential. In practice we minimized
$\int dt \left|n\left(t\right)-n_{KS}\left(t\right)\right|$
\cite{VerdozziPRL}, where $n_{KS}$ is the Kohn-Sham density, found by solving $i\dot{\psi}_{KS}=\left(\hat{t}+w+v_{KS}\right)\psi_{KS}$, and $\hat{t}$ is the kinetic term.

\noindent \textit{Ground state.} We start by solving the Dyson equation self consistently to obtain the initial $G$.
\begin{figure}[t]
\includegraphics[width=7.7cm, clip=true]{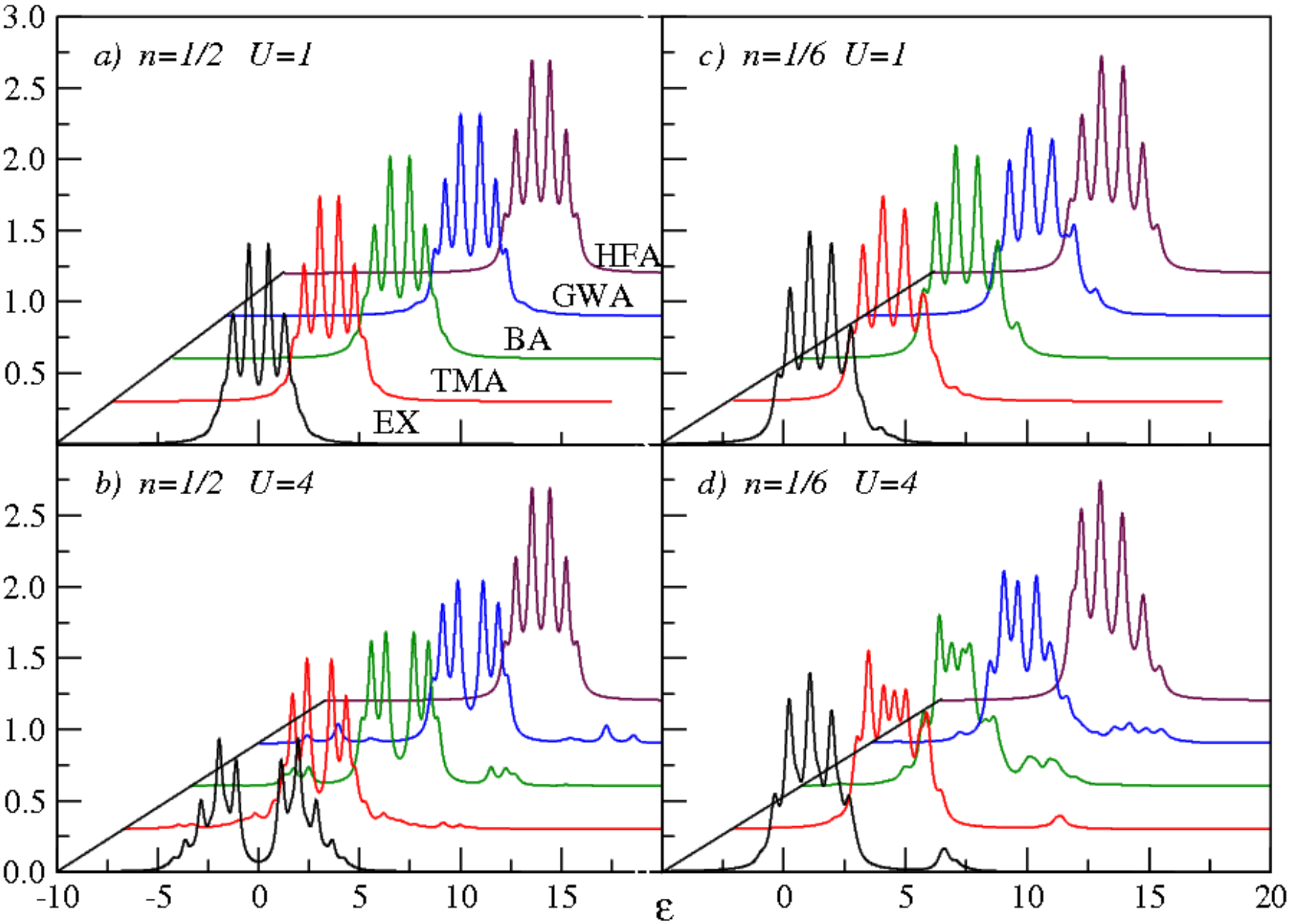}\\
\caption{ (Color online) Ground state spectral functions at $L$=6 for different fillings and interactions. 
The curves correspond to exact (black), TMA (red), BA (green), GWA (blue) and HFA (brown). The curves are shifted for clearer comparison and we have used a Lorentzian broadening $\Gamma=0.2V$.}
\label{gs}\end{figure}
Fig. \ref{gs} shows selected spectral densities for different particle concentrations and interactions. The case of half filling is shown in panels 2a) and 2b), where all the curves display the electron-hole symmetry. The exact solution exhibits the opening of a correlation gap on increasing $U$; this is also reproduced, at different extent, by the approximate treatments. The gap increase is most closely reproduced (but largely underestimated) by the BA, with the GWA and the TMA giving even smaller gaps. The exact lower and upper Hubbard bands are incorrectly reproduced by all the MBAs. In particular, for $U$=4, the MBAs introduce spurious satellite structures (less  pronounced in the TMA)  away from the band region. Results for the low density regime are in panels 2c) and 2d). For $U$=1, the main effect of the interaction is an increased asymmetry in the band region; the overall agreement between exact and MBAs is rather good, especially for the BA and the TMA. For $U$=4, the most notable feature in the exact solution is a satellite at about 6.5 (in an extended system, this would be a two-electron anti-bound state outside the continuum). Such satellite is well reproduced 
by the TMA, smeared out in the BA and the GWA, and obviously absent in the HFA. In the band region, for $U$=4, the agreement is moderate. 
We find that the spectral functions are better at first iteration.
This corresponds to the known fact that self-consistency often deteriorates the one-electron spectral properties. Self-consistency is required for dynamically conserving approximations and for total energies,  but other summation criteria should be adopted for spectral densities \cite{Almbladh}, and vertex corrections are often required to remove self-consistency artifacts \cite{MCCV87}. 
We found that the SGWA is slightly better than the GWA, since it  includes fewer 
faulty diagrams. Yet, it is still worse than the BA or the TMA. It is worth noting that  
the SGWA has a magnetic instability as a function of $U$. In the dimer, where the poles of $W[G_0]$ are $\epsilon=\pm\sqrt{4V^2 \pm 2VU}$, this occurs for $U \ge 2V$.
As a conclusive remark, self-consistent partial sums imply infinitely many diagrams, thereby violating the particle number constraint for finite systems. This results in an infinite, but discrete, number of poles in the spectral function. Even if this has no major effect in the ground state, it can have rather startling consequences for the TD behavior of the clusters.

\noindent \textit{Time dependence \Rmnum{1}: Densities. } The system is in the ground state for $t\le 0$.  At  $t>0$ an external field $w$ is applied. All results in the paper were obtained with a step-like $w\left(t\right)=w_{0}\Theta$$\left(t\right)$ (time unit = $|V|^{-1}$) applied only to the first ($R=1$) site, but we also examined other space and time dependencies.
\begin{figure}[t]
\includegraphics[width=7.7cm, clip=true]{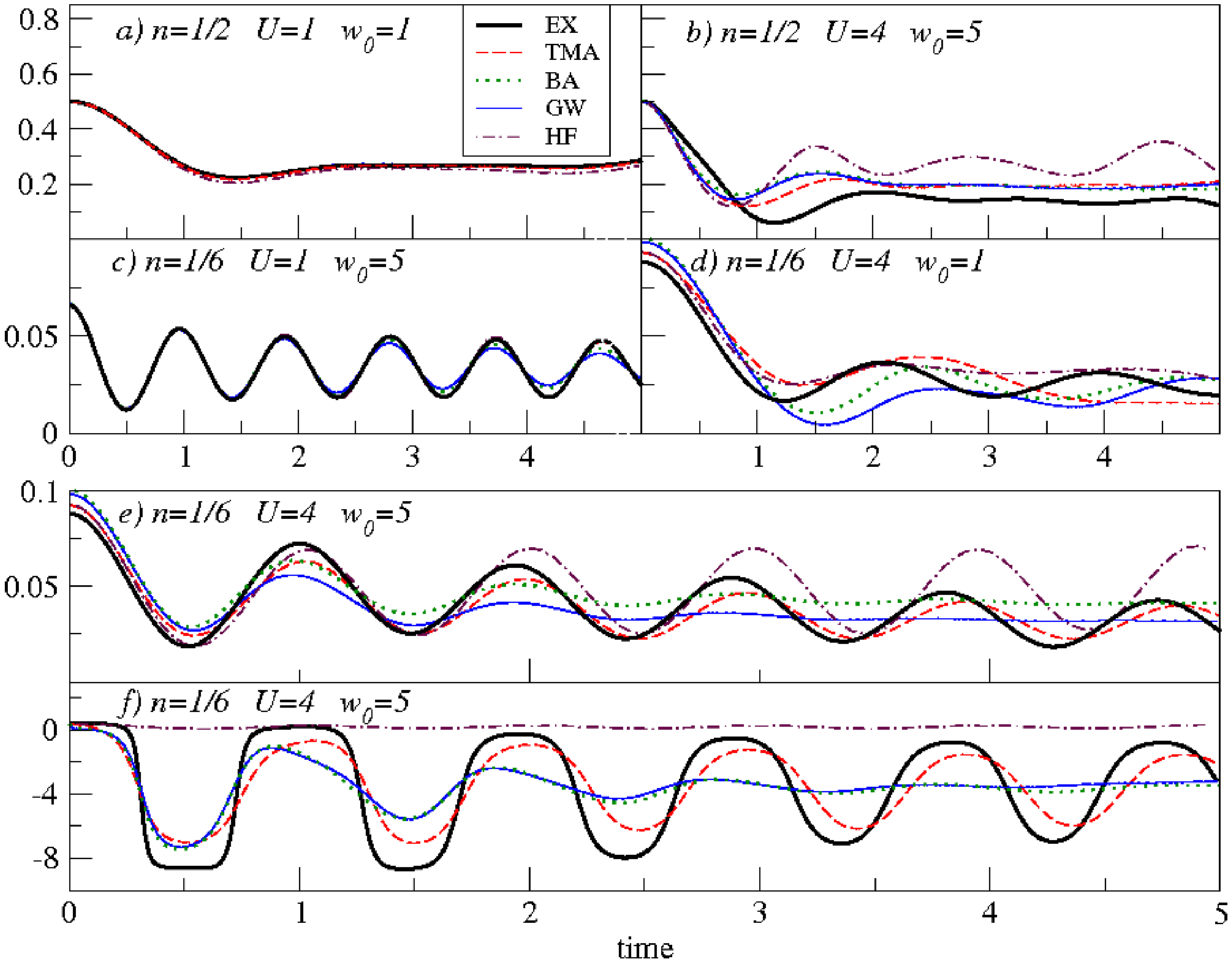}\\
\caption{ (Color online) TD densities (panels a to e) and $v_{KS}$ (panel f) on site 1. Exact (thick solid black), $TMA$ (dashed red), $BA$ (dotted green), $GWA$ (thin solid blue) and $HFA$ (brown dashed dot) results.}
\label{td}
\end{figure}
Fig. \ref{td} shows the TD  densities and $v_{KS}$  at $R=1$ for some representative cases.  
In panel a), where $n=1/2$ , we show the weakly interacting, weakly perturbed case $U=1, w_0=1$: All MBAs perform well. In panel b) both $U$ and $w$ are increased, but still $n=1/2$: the HFA performs poorly whilst the other MBAs are very similar to each other and closer to the exact density. In panel c), $n=1/6$, the HFA and TMA are virtually indistinguishable
from the exact density,  while the BA and GWA start to deviate after some time. In panel d), where $U=4, w=1$, none of the MBAs performs well. We attribute this to the poor
description of the ground state spectral function in the band region by the MBA's:
for $w=1$, the band region provides the main response to the perturbation. On the other hand, in panel e), the $TMA$ performs very well as the influence of the satellite becomes essential (the other MBAs either lack or completely misplace the satellite, see Fig. \ref{gs}). 
As a general comment, for short ranged interactions, as in our system(s), 
the TMA is rather suitable, especially at low densities. For the ground state, our results
are consistent with earlier studies \cite{MCCV87,CV95}. However, the novel and significant aspect 
here is that the performance in equilibrium of the MBAs  has considerable impact on
the TD behavior.  In particular, an external field redistributes the electrons to the unoccupied energy levels, and two spectral features, the energy gap and the satellite, play a key role in the TD evolution.  \newline
\noindent \textit{MBAs and TDDFT.} Our results also provide insight for TDDFT approaches to SCS 
(e.g., to the Hubbard model \cite{VerdozziPRL}). In Fig. 3, panel f), we show the KS
potentials obtained via reverse engineering from the MBAs and exact TD densities.
The performance of the different MBAs for $v_{KS}$ is consistent with the results for the TD densities.

\begin{figure}[t]
\includegraphics[width=7.7cm, clip=true]{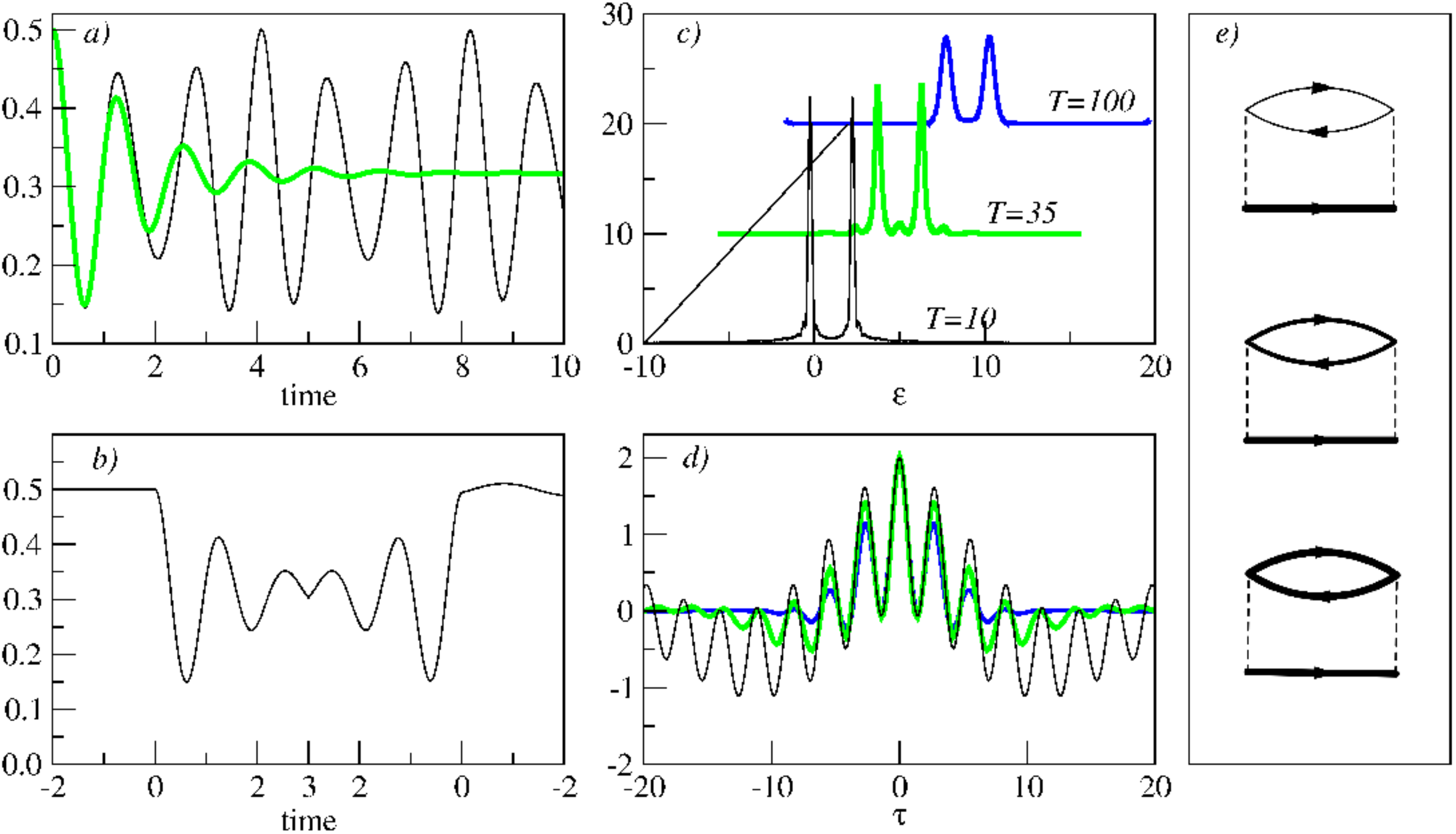}\\
\caption{ (Color online) $L=2, n=1, U=1$. a): Densities; exact (black) and damped $GWA$ (thick green) densities for $w_0=5$. b): Time reversed density. c): $A(T,\epsilon)$ for $w_0=2$ at $T=10$ (black), $T=35$ (thick green), $T=100$ (thick blue).  d): $A(T,\tau)$ for $w_0=2$ at $T=10$ (black), $T=35$ (thick green), $T=100$ (thick blue).}
\label{damping}
\end{figure}

\noindent \textit{Time dependence  \Rmnum{2}: Damping.} When we propagate the KBE in our approximate, self-consistent 
schemes we obtain damped solutions and strong numerical evidence of an 
(artificial) steady state (Fig. \ref{damping}a). The damped dynamics fulfills
time-reversal symmetry (Fig. \ref{damping}b). The damping increases with the strength of the external TD  field, 
and is absent in the linear response limit \cite{BetheSalp}.
We investigated  the instantaneous spectral function
$A\left(T,\omega\right)=-Tr\: Im\int_{-2T}^{2T} e^{i\omega \tau}\left[G^{>}-G^{<}\right]\left(T+\frac{\tau}{2},T-\frac{\tau}{2}\right)d\tau$ and its counterpart in time space, where $T=\left(t_{1}+t_{2}\right)/2$ and $\tau=\left(t_{1}-t_{2}\right)$.
At the steady state, $A$
gets broadened in energy space (Fig. \ref{damping}c) and damped in time space 
(Fig. \ref{damping}d). The time required to reach the steady state depends on the MBA used - in general the TMA is slowest - and it generally
increases with system size. The damping is largest at the perturbed site. The steady states also depends on how the perturbation
is switched on. With a sudden switch-on we reach a steady state with the
same energy as the initial ($t=0^+$) state. With an increasingly slow switch-on, 
the steady state gets progressively closer to the ground-state with
a static perturbation, in line with the adiabatic theorem. We note that each steady state in a TD scheme corresponds to
a solution of the steady KBE. Thus, in our MBAs, the steady KBE have multiple solutions. 
Some of these are artificial,  they do not correspond to thermal equilibrium and most probably have
continuous spectral functions. 

Artificial damping in finite systems does not occur in wavefunction-based treatments. 
However, self consistent MBAs are entirely
defined by a generating functional $\Phi[G]$ \cite{KB-book} and a corresponding
set of self-consistent KBE: there is no direct connection to wavefunctions.
In finite systems, MBAs involving partial infinite-order
summations include diagrams with 
which annihilate more particles than present in the system.
In an "exact'' theory, these unphysical contributions will give
no net contribution, but such perfect cancellation is in general
lost in approximations like BA, GWA or TMA.
Our numerical results show that these MBAs give ground state spectral functions
with an incorrect pole structure, with discrete but infinitely many poles. 
In the time dynamics, these unphysical aspects of our MBAs have more drastic consequences. 
To be specific, we discuss three particle-conserving versions of the BA with increasing level of self-consistency \cite{Ulf}. 
If we evaluate the polarization $P$ with ground-state propagators (Fig. \ref{damping}e top),
the solution does not damp. If we instead evaluate $P$
with propagators in the TD HFA  (Fig. \ref{damping}e middle), the 
solution is partially damped. Finally, with
a fully self-consistent $P$ (Fig. \ref{damping}e bottom), damping occurs. 
Thus, if all $G$s in 
$\Sigma$ vary consistently with the external
field, we have the remarkable result that a finite system reaches an (artificial) steady state \cite{except}.

In conclusion, we studied the non equilibrium dynamics of clusters with strong
correlations, by comparing exact results to those from four many-body approximations (MBAs).
Among such MBAs, the T-matrix approximation performs very
well at low densities and is in general superior to the GW and 2nd Born
schemes. Our results also show the principle limitations of conventional MBAs for finite systems,
since we obtained damped solutions and artificial steady states. We attribute 
this to the inconsistency of applying infinite-order 
but approximate summations to finite systems. We acknowledge Robert van Leeuwen for fruitful 
discussions and Ulf von Barth for critical reading. This work was supported by the 
EU 6th framework Network of Excellence NANOQUANTA (NMP4-CT-2004-500198) and the 
European Theoretical Spectroscopy Facility (INFRA-2007-211956).


\begin{thebibliography}{99}
\bibitem{BetheSalp}
In this limit the response is described by the Bethe-Salpeter equation, with a kernel
$\delta \Sigma / \delta G$. The latter would have a discrete spectrum  in our MBAs, and so would the resulting density response.
\bibitem{except} This conclusion applies to all MBAs here, although partially self-consistent,
but particle-conserving schemes can be easily devised for GWA but not for TMA.
\end{thebibliography}
\end{document}